\documentstyle[floats,aps]{revtex}

\begin{document}
\draft

\catcode`\@=11 \catcode`\@=12 \twocolumn %
[\hsize\textwidth\columnwidth\hsize\csname@twocolumnfalse\endcsname
\title
{Effect of Subband Landau Level Coupling to the Linearly
Dispersing Collective Mode in a Quantum Hall Ferromagnet}
\author{Yue Yu}
\address{Institute of Theoretical Physics, Chinese Academy of Sciences, P. O. Box 2735,
Beijing 100080, China}

\date{\today}
\vspace{0.1in}

\maketitle
\begin{abstract}
In a recent experiment (Phys. Rev. Lett. {\bf 87}, 036803 (2001)),
Spielman et al observed a linearly dispersing collective mode in
quantum Hall ferromagnet. While it qualitatively agrees with the
Goldstone mode dispersion at small wave vector, the experimental
mode velocity is slower than that calculated by previous theories
by a factor of about 0.55. A quantitative correction may be
achieved by taking the subband Landau level coupling into account
due to the finiteness of the layer thickness, which gives a better
agreement with the experimental data. A method coupling the
quantum fluctuation to the tunneling is briefly discussed.
\end{abstract}

\pacs{PACS numbers:  73.43.-f,71.35.Lk,73.21.-b,73.40.Gk} ]

The fractional quantum Hall effects as well as most of related
phenomena reflect the behavior of the ground states of the
two-dimensional electron gas in high magnetic field at zero
temperature \cite{q1,q2}. However, a first experimental signal of
the finite temperature phase transition in such  systems has been
recently observed by Spielman et al in a bilayer system for
$\nu_T=1$ \cite{spie1,spie2}. This transition seems to be closely
related to a Kosterlitz-Thouless transition into an earlier
predicted broken symmetry state \cite{wen} which was akin to the
Josephson tunneling in superconductivity. However, differing from
the Josephson effect in which the zero-bias conductance at the
zero temperature was divergent, the conductance peak in that
experiment was finite even extrapolating to the zero-temperature.
This phenomenon, with other intriguing novel properties, caused a
set of theoretical research works
\cite{sch,stern,bal,stern1,fog,kim,jog,vei,buk,yueyu} in which
various scenarios have been suggested while there is still a
variety of open issues ( for a short review, see \cite{gir}).

Among these issues, we shall focus on the finite layer thickness
affecting the tunneling through the tilted field, which was
disregarded by literature. As a result from the zero thickness
approximation, a quantitative discrepancy between the previous
theoretical calculation and the experimental data already
appeared. The precise description of the issue is as follows: In a
recent experiment \cite{spie2}, Spielman et al observed a linearly
dispersing collective mode in a small wave vector in bilayer
two-dimensional electron systems, which was identified as the
pseudospin Goldstone mode long expected. However, Figure 4 of
their published paper showed the sound velocity, namely, the slope
of the linear dispersion experimentally observed was smaller than
that theoretically predicted by a factor of about 0.55. One argued
that this discrepancy may result from the overestimate in the
Hartree-Fock approximation used in the theoretical calculation,
and the quantum fluctuation may repair this discrepancy. However,
the present exact diagonalization result indicated that the
correction from the quantum fluctuation does not seems to reach
such a small factor \cite{moon}. Thereby, it is worth to look for
other sources to influence the mode.

On the other hand, due to the dispersion proportional to the
in-plane magnetic field, the finite thickness of the layers may
affect it because the layer thickness is comparable to the
interlayer spacing in the experiment. However, the previous
theoretical treatments were essentially based on the zero-layer-
thickness approximation. The experiences in the study of the
quantum Hall systems were that finite thickness effects often
determine the quantitative consistence between the theory and
experiment, e.g., the gap of Laughlin's state, and so on.

Moreover, for this bilayer system, the ground-state behavior is
not well understood yet if there is a tilted field and the finite
layer thickness is taken into account. We even do not have a
satisfactory variational ground-state wave function. Thus, one may
suspect whether this experimental observed mode can be identified
as the theoretically anticipated pseudospin Goldstone mode. In
this paper, we deal with the experimental data through a
theoretical model which attributes the finite thickness correction
to the sound velocity. If the correction is positive, we may say
that the identification can be accepted and otherwise it could be
more suspect. In a composite boson picture when the field is
tilted \cite{Yu}, while the composite bosons see an opposite
effective perpendicular field with an equal magnitude in the mean-
field state, they see a weakened effective parallel field
$B^*_\parallel$ due to the finite layer thickness via the subband
Landau-level coupling. Thus, one can apply the the interlayer
tunneling theory \cite{bal,stern,fog} to this composite boson
system but the parallel field $B_\parallel$ is reduced to
$B^*_\parallel$. By using the Hartree-Fock estimate of the sound
velocity \cite{gm,gir}, one can obtain the linear dispersing
Goldstone mode in the same way as the theoretical line in Fig. 4
of \cite{spie2} but $q$ is replaced by $q^*=eB^*_\parallel
d/\hbar$. This dispersion has a substantial improvement to fit the
experimental data. Thus, we have a positive correction. In the
meanwhile, our composite boson formalism shows there is a coupling
of the quantum fluctuation to the tunneling. However, a detailed
discussion of the quantum fluctuation has exceeded the goal of the
present work.

In this work, we only deal with the finite-thickness effect via
the subband Landau-level coupling and neglect the others, say,
affecting the Coulomb interaction due to the finite thickness. In
order to deal with the subband Landau-level coupling analytically,
we assume that the electron gas is confined in a plane by an
infinite harmonic potential. Before going to the details, we argue
that this choice of the confining potential may quantitatively
correct the theoretical calculation without severely impacting the
comparability between our calculation and the experimental data
despite the fact that the realistic confining potential in the
sample is essentially a finite square well. The harmonic well is
very different from the square well in their excited spectra, for
the harmonic spectrum is equal gaped while that of the square well
is not. However, the temperature is extremely low so that only the
lowest subband of the lowest Landau level is filled in the present
situation. Thus, no excited spectra will be involved. In the
ground state, one may variationally adjust the harmonic frequency
such that the ground state wave function has the best shape, and
let the subband energy equal that in a more realistic square well.
In this sense, the harmonic potential may be a good approximation
to a realistic potential, to give a quantitative correction due to
the subband coupling. Such a harmonic potential has been chosen to
deal with many quantum Hall systems to replace the realistic
potential which is either triangular \cite{cha} or square
\cite{mac}. It was also used to discuss a giant magnetoresistance
induced by a parallel magnetic field \cite{das}.

We start from the problem of a single particle in a strong
magnetic field which is tilted at an angle to the x-y plane. An
in-plane field in the $x$-direction violates the two-dimensional
rotational symmetry. By introducing a harmonic confining potential
with the character frequency $ \Omega$ in the $z$-direction, the
electron is restricted to quasi-two-dimensions. The
single-particle
Hamiltonian can be diagonalized as $H_{{\rm s.p.}}=\hbar\omega_-\alpha^%
\dagger_\xi\alpha_\xi+\hbar\omega_ +\alpha^\dagger_z\alpha_z$ with
the diagonalized oscillators given by
$
\alpha_\xi^\dagger=(a_\xi^\dagger,a_z^\dagger,a_\xi,a_z)X^{-T},
\alpha_z^\dagger=(a_\xi^\dagger,a_z^\dagger,a_\xi,a_z)X^{+T} $,
where $a^\dagger_\xi=\frac{1}{\sqrt{2}}(-\partial_\xi
+\frac{1}{2}\bar\xi),a_\xi=\frac{1}{\sqrt{2}}(\partial_{\bar\xi}
+\frac{1}{2}\xi)$ and
$a^\dagger_z=\frac{1}{\sqrt{2}}(-\frac{\partial}{\partial
z'}+z'),a_z=\frac{1}{\sqrt{2}}(\frac{\partial}{\partial z'}+z')$.
$z'=\hat\Omega^{1/2} z$. The vectors $X^\pm$ are given by
\begin{eqnarray}
X^-&\propto&(\omega_c+\omega_-,
-\frac{\tilde\omega\omega_c}{\tilde\Omega-\omega_-},
-(\omega_c-\omega_-),\frac{\tilde\omega\omega_c}{\tilde\Omega+\omega_-})
,\nonumber\\ X^+&\propto&
(-\frac{\tilde\omega\tilde\Omega}{\omega_c-\omega_+},
\tilde\Omega+\omega^+,\frac{\tilde\omega\tilde\Omega}{\omega_c+\omega_+},
-(\tilde\Omega-\omega^+)).
\end{eqnarray}
The frequencies $\omega_\pm$ are given by
$
\omega_\pm^2=\frac{1}{2}(\tilde\Omega^2+\omega_c^2)\pm\frac{1}{2} \sqrt{(%
\tilde\Omega^2-\omega_c^2)^2+4|\tilde\omega|^2\tilde\Omega\omega_c},
$
where $\tilde\omega=\omega_x(\omega_c/\tilde\Omega)^{1/2}$ and $\tilde\Omega%
^2=\Omega^2+\omega_x^2$; $\omega_x$ and $\omega_c$ are the
cyclotron frequencies corresponding to $B_x$ and $B_z$. Here we
have applied the unit $l_B=\sqrt{\hbar c/eB_z}=1$. In addition,
there is a conservation quantity $L_\xi=
\tilde a_L^\dagger\tilde a_L$ with $\tilde a_L=\frac{1}{\sqrt{2}}%
(\partial_\xi+\frac{1}{2}\bar\xi)$ and $\tilde
a^\dagger_L=\frac{1}{\sqrt{2}}
(-\partial_{\bar\xi}+\frac{1}{2}\xi)$ irrespective of whether the
tilted angle $\theta=0$ or not. To solve this single-particle
problem, we seek the ground state which is the eigenfunction of
$L_\xi$. It
is useful to make a coordinate rotation with $\xi\to\tilde\xi=\xi+\beta\bar\xi%
+\gamma z^{\prime}$ and $\tilde z^{\prime}=z^{\prime}$ with
$\beta$ and $\gamma$ determined by $[\alpha_\xi, \tilde\xi]=
[\alpha_z, \tilde\xi]=0$:
\begin{eqnarray}
\beta&=&\frac{(X_2^+-X_4^+)X^-_3-(X_2^--X_4^-)X^+_3}
{(X_2^+-X_4^+)X^-_1-(X_2^--X_4^-)X^+_1},\nonumber\\
\gamma&=&\frac{2(X^+_3X^-_1-X^-_3X^+_1)}
{(X_2^+-X_4^+)X^-_1-(X_2^--X_4^-)X^+_1}.
\end{eqnarray}
The ground state wave functions are highly degenerate and of the
form $\Psi_0(\tilde\xi,\tilde\xi ^*,\tilde
z^{\prime})=f(\tilde\xi)e^g$ with $g(\tilde\xi,\tilde\xi ^*,\tilde
z^{\prime})$ being a quadratic form of $\tilde\xi,\tilde\xi ^*,\tilde z%
^{\prime}$ whose coefficients are determined by $\alpha_\xi
e^g=\alpha_z e^g=0$. The function $f(\tilde\xi)$ is an arbitrary function of $\tilde%
\xi$.

Notice that linear-independent wave functions $\tilde\xi^m e^g$
(m=0,1,2,...) are not the eigen functions of $L_\xi$. However, one
can start from those linear-independent wave functions to
construct the common eigen functions of $H_{{\rm s.p.}}$ and
$L_\xi$, which read $f_m(\tilde\xi)e^g,$ with $f_m(\tilde\xi)
=\sum_{m^{\prime}=0}^{M-1}f_{mm^{\prime}} \tilde\xi^m$ for $M$
being the number of Landau orbits and $M=N$ for $\nu_T=1$. The
coefficients $f_{mm^{\prime}}$
are dependent on the in-plane field and confined by $f_{mm}(0)=1$ and $%
f_{mm^{\prime}}(0)=0$ for $m\ne m^{\prime}$ if $\theta=0$. Those
degenerate ground state wave functions are orthogonal and with the
eigen value $m$ of $L_\xi$.

After solving the single-particle problem, we turn to the
many-body ground state wave function. The common Laughlin's state
or Halperin's (111)-state for the vanishing tilted angle is no
longer a good variational wave function. However, to be
enlightened by them, we postulate the many-body ground states for
$\nu_T=1$ as $$\Psi_0(\vec r_1,...,\vec r_N)={\rm St}
(f_0,...f_{N-1})\exp(\sum_i g_i)\times |{\rm PSS}\rangle,$$ where
$\vec r_i$ are the three-dimensional
position vectors and ${\rm St}$ is the Slater determinant of $%
f_0(\tilde\xi_{P(1)}),..., f_{N-1}(\tilde\xi_{P(N)})$. $|{\rm
PSS}\rangle$ denotes the possible layer coherence, which is either
pseudospin fully polarized or coherent. Although the quantity
$L=\sum_i L_{\xi_i}$ is no longer conserved due to the
interaction, there is a conserved quantity $L+\sum_{i\ne j}\tilde
a_{\xi_i}^\dagger\tilde a_{\xi_j}$. The constructed state is the
eigenstate of this quantity with  eigen value 0. Due to $M=N$, one
has exactly
\begin{eqnarray}
\Psi_0(\vec r_1,...,\vec
r_N)\propto\prod_{i<j}(\tilde\xi_i-\tilde\xi_j) \exp(\sum_i
g_i)\times|{\rm PSS}\rangle,\label{wf}
\end{eqnarray}
where, for the balanced bilayer system, $i=1,...,N/2$ and
$1+N/2,...,N$ denote the particles in layers 1 and 2,
respectively. The spatial part of this wave function goes back to
Halperin's (111)-state for $B_\parallel=0$. Corresponding to the
system in the experiment of Spielman et al \cite{spie1,spie2}, the
coherent state is given by
\begin{eqnarray}
&&|{\rm PSS}\rangle=|\to\to...\to\rangle,\\
&&|\to\rangle=\frac{1}{\sqrt
2}(|\uparrow\rangle+e^{i\varphi}|\downarrow\rangle),\nonumber
\end{eqnarray}
if one uses $|\uparrow,\downarrow\rangle$ to represent the
electron in the upper or lower layer.

Since the electrons in the quantum Hall state are strongly
correlated, when an electron tunnels from one layer to another,
the fluxes combined with the electron also move accompanying the
electron. If the magnetic field is not tilted, the fluxes
accompanying the electron do not move in the tunneling because
they are perpendicular to the x-y plane. However, if the field is
tilted, the fluxes accompanying the electron no longer lie in the
x-y plane due to $\xi\to\tilde \xi$. Thus, the electron tunneling
causes the flux hopping from one layer to another. This can not be
reflected in the single electron tunneling picture. A better
formalism is the composite boson formalism. The single composite
boson tunneling counts the charge tunneling and the flux hopping
simultaneously. Hence, we use the composite boson formalism.
According to (\ref{wf}), for a $\nu_T=1$ state, the composite
boson theory in a tilted magnetic field can be achieved by the
anyon transformation in the spatial wave function \cite{Yu}
\begin{eqnarray}
\Psi(\vec r_1,...,\vec r_N)=\Pi_{i<j} \frac{\tilde\xi_{ij}}
{|\tilde\xi _{ij}|} \Phi(\vec r_1,...,\vec r_N).
\end{eqnarray}
In terms of Ref.\cite{Yu}, this transformation gives a statistical
gauge field, $ a_\mu(\vec r_i)=-\sum_{j\not{=}i}
\frac{\tilde\epsilon _{\mu\nu} \tilde
x^{\nu}_{ij}}{|\tilde\xi_{ij}|^2},a_z(\vec r_i)
=-\frac{\tilde\Omega^{1/2}}{\omega_c^{1/2}} \sum_{j\not{=}i}
\frac{\gamma\tilde y_{ij}} {|\tilde \xi_{ij}|^2},
$
where $\tilde\xi= \tilde x+i\tilde y$;
$\tilde\epsilon_{12}=1+\beta$ and $\tilde \epsilon_{21}=-1+\beta$.
The corresponding statistical magnetic field $\vec
b=\nabla\times\vec a$, i.e.,
\begin{eqnarray}
b_z(\vec\xi_i)&=&2\pi\sum_{j\not{=}i}\delta^{(2)}(\xi_{ij}),\nonumber\\
b_\parallel(\vec\xi_i)&=&-2\pi(1+\beta)^{-1}(\tilde\Omega/\omega_c)^{1/2}
\gamma\sum_{j\not{=}i}\delta^{(2)}(\xi_{ij}).
\end{eqnarray}
In the mean field approximation, $\bar b_z=2B_z$, while $\bar
b_\parallel$ partially cancels $B_\parallel$. The effective
parallel magnetic field seen by the composite boson reads
\begin{eqnarray}
B^*_\parallel=(1-(\gamma/1+\beta)\sqrt{\tilde\Omega\omega_c}
/\omega_\parallel)B_\parallel\equiv DB_\parallel.
\end{eqnarray}
We shall see that $D\leq 1$ and the equality holds if
$\omega_c/\Omega\to 0$. Since $\gamma/(1+\beta)\propto
\frac{\omega_\parallel}{\omega_c}+O(\frac{\omega_\parallel^3}
{\omega_c^3})$ for $\omega_\parallel\ll\omega_c$, $D$ is almost
independent of $B_\parallel$ for the parameters used in
\cite{spie2}. The inter-layer tunneling operator reads
\begin{eqnarray}
T&=&-t\int dx
dy\psi^\dagger_\uparrow(x,y)e^{iA_zd}\psi_\downarrow(x,y)+h.c.\nonumber\\
&=&-t\int dxdy
\phi^\dagger_\uparrow(x,y)e^{i(A_z-a_z)d}\phi_\downarrow(x,y)+h.c.,
\label{full}
\end{eqnarray}
where $\psi^\dagger_\uparrow$ and $\phi^\dagger_\uparrow$ denote
the electron and composite boson creation operators in the upper
layer respectively,etc. $d$ is the interlayer spacing. $A_z$ is
the the $z$-axial component of the vector potential of the
external magnetic field and $a_z$ is that corresponding to $\vec
b(\vec r)$. Thus, in the mean field approximation,
$\phi_{\uparrow,\downarrow}=\sqrt{\rho_0}e^{i\theta_{\uparrow,\downarrow}}$
($\varphi=\theta_{\uparrow}-\theta_{\downarrow}$),
$A_z-a_z=DB_\parallel x $ and the tunneling Hamiltonian reads
\begin{eqnarray}
T=-\int dxdy \frac{t}{2\pi}\cos(\varphi-q^*x),\label{mf}
\end{eqnarray}
where the effective wave vector $q^*=Dq$ instead of $q=2\pi
B_\parallel d/\Phi_0$. Thus, the theory \cite{moon} applied to the
electron can also apply to the composite boson because the anyon
transformation is carried out in the spatial wave function and
does not change the particle density and the pseudospin coherence
state. The only difference in the final result from that in the
electron's case is replacing $q$ by $q^*$. That is, for small
$q^*$, the linear dispersion is given by $vq^*$. However, all
measurements in the experiment by Spielman et al \cite{spie2}
correspond to $B_\parallel$, namely, $q$. To compare with the
experimental data, the dispersion may written as
\begin{eqnarray}
vq^*=Dvq=v^*q.
\end{eqnarray}
Due to $D\leq 1$, $v^*$ is smaller than $v$ by a factor $D$. If we
use the Hartree-Fock value of the sound velocity to estimate $v$,
the dispersion is exactly what is plotted in Fig. 4 of
\cite{spie2} except that $q$ is replaced by $q*$; or in the wave
vector $q$, the sound velocity $v$ is renormalized
non-perturbatively to $v^*$. This renormalization is remarkably
different from that by the quantum fluctuation in the
zero-thickness theory and is the result of the strong correlation
of the electrons.

It is easy to see that the magnitude of $D$ is determined by the
ratios $\omega_\parallel/\omega_c$ and $\omega_c/\Omega$. Since
the tunneling peak is destroyed in a small tilted angle, say
$B_\parallel\sim 0.6$ T (thus the critical value of the ratio is
$\omega_\parallel/\omega_c\sim 0.3 $ for $n_T=5.2\times
10^{10}/$cm$^2$ because $B_z\sim 2$T ) \cite{spie2}, one can
consider the small ratio $\omega_\parallel/\omega_c$ only.  In
fact, one finds that $D$ is almost independent of $B_\parallel$ if
$\omega_\parallel/\omega_c<0.2$. In Table 1, we list the
magnitudes of $D$ for $\omega_\parallel/\omega_c=0.2$.

\vspace{0.1in}

\begin{tabular}{ccccccccc}
\hline\hline
$~\frac{\omega_c}{\Omega}~~$&~0.10~&~0.40~&~0.50~&~0.80~&~1.00~&~1.25~&1.67~
&2.50 \\ \hline
D~~&~1.00~&~0.90~&~0.86~&~0.76~&~0.70~&~0.65~&~0.60~&0.55\\
\hline\hline
\end{tabular}

\vspace{0.1in}

{\small Table 1 The magnitudes of $D$ for the different ratios
$\frac{\omega_c}{\Omega}$

and $\frac{\omega_\parallel}{\omega_c}=0.2$.}

\vspace{0.1in}

The first column in Table 1 implies that it is back to the result
got by ignoring the finite thickness. The last value of $D$
reaches the experimental data but the corresponding ratio
$\frac{\omega_c}{\Omega}=2.5$ is too large for the sample. In the
experiment sample, the square well has the width $a=18$nm. The
magnetic length $l_B\sim 17$nm for the density $n_T=5.2\times
10^{10}$ and $\nu_T=1$. The ratio $\Omega/\omega_c$ is dependent
on the heights of the square well. For the potentials with the
infinite height on two sides of the well, $\omega_c/\Omega\sim
\frac{a^2}{\pi^2 l_B^2}\sim 0.11$ and $D=0.99$ if we determine
$\Omega$ by $\frac{1}{2}\hbar\Omega\sim
E_0=\frac{\pi^2\hbar^2}{2ma^2}$. If this was the case, there would
be almost no finite thickness effect. However, the wells in the
real bilayer systems are finite and asymmetric. The single
particle ground state energy may be lowered substantially. If the
potential $V_1$ on one side is lower than $V_2$ on the other side,
the ratio is given by $$\omega_c/\Omega\sim (kl_B)^{-2},$$ where
$k\leq k_1$ is determined by the solution of the equation
$ka=\pi-\sin^{-1}(k/k_1)-\sin^{-1}(k/k_2)$ with $k_i=\sqrt
{2mV_i/\hbar}$. Thus, the ratio is dependent on the heights of the
well on both sides. Assuming $V_1=V_2$ and for a typical well with
$V_1\sim2\hbar^2/(ma^2)$, the value of $k^2$ may reduce to
$0.15\pi^2/a^2$. Hence, $$\omega_c/\Omega\sim
a^2/(0.15\pi^2l_B^2)\sim 0.7~~~{\rm and}~~~D\sim 0.78.$$ The
details of the well heights were not reported in
\cite{spie1,spie2}. We do not expect that the result stemming from
the mean field approximation can completely fit with the
experiment. However, what we can say is that the finite thickness
correction drives the theoretical results to a positive direction
to be consistent with the experimental data. There may be other
sources to improve the theoretical result. Especially, the quantum
fluctuations may be very important because it lowers the spin
stiffness and then flatten the slope. Typically, the quantum
fluctuation may lower the stiffness by a factor $\leq 0.80$ and
then a factor $\leq 0.89$ to the sound velocity according to an
exact diagonalization result by Moon et al \cite{moon}. Combining
this factor with the factor from the finite thickness correction,
the theoretical result may have a good agreement with the
experimental data.

Before finishing this paper, we would like to point out that the
composite boson formalism we used here introduces a coupling of
the phase $\varphi$ to the quantum fluctuation other than that
discussed in literature \cite{moon}. Note that from (\ref{full})
to (\ref{mf}), we use the mean field approximation. That is,
$\delta a_z$ in $A_z-a_z=\bar A_z-\delta a_z$ has been neglected.
The gauge fluctuation couples to the phase in the tunnel
Hamiltonian is given by
\begin{eqnarray}
T&=&-\frac{t}{2\pi} [\cos(\varphi-q^*x)\cos(\delta
a_zd)\nonumber\\ &-&\sin(\varphi-q^*x)\sin(\delta a_zd)].
\end{eqnarray}
Obviously, the gauge fluctuation will weaken the tunneling, then
the charged gap $\Delta_{\rm SAS}$. We will leave further
discussion of this aspect to a separate work \cite{note}.

In conclusion, we have considered the subband Landau level
coupling in the interlayer tunneling of the bilayer quantum Hall
system. It is found that this coupling may quantitatively affect
the linear dispersion of the pseudospin Goldstone mode. A further
improvement to the linear dispersion requires more sample
parameters, especially, the heights of the confining potential as
well as to develop a reliable method to count the correction from
the quantum fluctuation.

This work was supported in part by the NSF of China.


\begin{references}
\bibitem{note} For example, we can discuss the commensurate-incommensurate
phase transition \cite{muphy} investigated in \cite{moon}. The
weaker tunneling, the weaker critical parallel field. Then, our
critical field $B_{\parallel,c}$ differs from that in \cite{moon}
a factor about $\sqrt{\langle\cos \delta a_z d\rangle}/D$.
\end{references}
\end{document}